\documentclass[aps,twocolumn,nofootinbib,preprintnumbers,groupedaddress,letterpaper]{revtex4}
\usepackage[dvips]{color}
\usepackage[normalem]{ulem}
\usepackage{amsmath}
\usepackage{enumerate}
\usepackage{amsfonts}
\usepackage{epsfig}
\usepackage{yfonts}

\usepackage{graphicx}
\usepackage{graphicx}
\usepackage{bm}
\usepackage{subfigure}

\newcommand\comment[1]{}

\newcommand\poincare{Poincar\' e }

\def\le{\left}
\def\ri{\right}

\def\({\left(}
\def\){\right)}
\def\[{\left[}
\def\]{\right]}
\def\<{\langle}
\def\>{\rangle}
\newcommand\half{{\ensuremath{\frac{1}{2}}}}
\newcommand\p{\ensuremath{\partial}}

\newcommand\field[1]{{\ensuremath{\mathbb{{#1}}}}}

\newcommand{\RR}{\field{R}}

\newcommand{\be}{\begin{equation}}
\newcommand{\ee}{\end{equation}}
\newcommand{\bea}{\begin{eqnarray}}
\newcommand{\eea}{\end{eqnarray}}
\newcommand{\bwt}{\begin{widetext}}
\newcommand{\ewt}{\end{widetext}}

\newcommand{\bi}{\begin{itemize}}
\newcommand{\ei}{\end{itemize}}
\newcommand{\ben}{\begin{enumerate}}
\newcommand{\een}{\end{enumerate}}
\newcommand{\bca}{\begin{cases}}
\newcommand{\eca}{\end{cases}}
\newcommand{\bln}{\begin{align}}
\newcommand{\eln}{\end{align}}
\newcommand{\bst}{\begin{split}}
\newcommand{\est}{\end{split}}

\begin{document}

\preprint{CERN-PH-TH-2015-213 }

\title{Colliding waves on a string in AdS$_3$}

\author{David Vegh}
\email{dvegh@cmsa.fas.harvard.edu}

%School of Natural Sciences,

\affiliation{\it  CMSA, Harvard University, Cambridge, MA 02138, USA  \\ }
\affiliation{\it  Institute for Advanced Study, Princeton, NJ 08540, USA \\  }
\affiliation{\it Theory Group, Physics Department, CERN, CH-1211 Geneva 23, Switzerland}

\date{\today}

\begin{abstract}

This paper is concerned with the classical motion of a string in global AdS$_3$. The initially static string stretches between two antipodal points on the boundary circle. Both endpoints are perturbed which creates cusps at a steady rate. The cusps propagate towards the interior where they collide.
The behavior of the string depends on the strength of forcing. Three qualitatively different phases can be distinguished: {\it transparent, gray}, and {\it black}. The transparent region is analogous to a standing wave. In the black phase, there is a horizon on the worldsheet and cusps never reach the other endpoint. The string keeps folding and its length grows linearly  over time. In the gray phase, the string still grows linearly. However, cusps do cross to the other side.
The transparent and gray regions are separated by a transition point where a logarithmic accumulation of cusps is numerically observed. A simple model reproduces the qualitative behavior of the string in the three phases.

\end{abstract}

\maketitle

\section{Introduction}

This paper investigates the formation of worldsheet horizons on a long string that moves in a fixed background geometry. If one perturbs the string endpoint, then a wave is created that propagates down the string. At the location of the wavepacket, the string is effectively longer now. Any perturbation that travels through this region suffers a time delay\footnote{This is reminiscent of the Shapiro time delay of general relativity.}. Event horizons form when the rate of waves is large enough so that perturbations cannot cross them. Here we would like to understand the transition from a horizon-free string to one with an event horizon as  the amplitude of waves crosses a critical value.

The nature of the problem requires the study of steady states where the initial transient oscillations have already died out. This demands very stable numerical calculations. An exact discretization technique \cite{Vegh:2015ska} (see also \cite{Callebaut:2015fsa}) will be used that operates by adding elementary shockwaves (i.e. cusps) to the string.
We will consider a string in (2+1)-dimensional global anti-de Sitter spacetime. The string stretches between two points on the boundary which is a circle. According to the AdS/CFT correspondence \cite{Maldacena:1997re, Gubser:1998bc, Witten:1998qj}, the boundary gauge theory contains a Wilson loop on which the string ends \cite{Maldacena:1998im, Rey:1998ik}. The Wilson loop lies along the path of an infinitely heavy quark-antiquark pair and the string is the holographic dual to the color flux tube that connects them.

On the \poincare patch, the building blocks of \cite{Vegh:2015ska} are shrinking/expanding circular strings that all have worldsheet horizons. Since we would like to start with a horizon-free string, here {\it global} AdS will be considered\footnote{  The construction of a static string that stretches between two boundary points \cite{Hubeny:2014zna} presumably involves an infinite number of cusps. The waves created by perturbing the endpoints would propagate in this bath of cusps. See \cite{Ishii:2015wua} for related numerical simulations.}. Both endpoints of the string will be perturbed. This creates (shock)waves on the string that propagate with the speed of light.
The quark and the antiquark will be kicked\footnote{By a slight abuse of language, discontinuous jumps in the acceleration of the string endpoint will be referred to as ``kicks''. In the absence of kicks, the string endpoint moves with a constant acceleration on the \poincare patch.} in alternating direction at equal time spacing starting at $t=0$. After some time, this non-equilibrium system reaches a steady state. Depending on the rate and the strength of the kicks, the string can move in qualitatively different ways:
\begin{itemize}
\item
For small kicks (not necessarily in the linear regime), the cusps on the string pass through each other and a standing wave forms. The string length oscillates around a constant value.
\item
For large enough kicks, an event horizon forms on the string worldsheet. Even though individual cusps pass through each other, a cusp might not make it all the way across the string, since on the other side new cusps are constantly being added. The string grows linearly in time.
\item
Surprisingly,  for intermediate kick strengths, there is another phase. Even though the string has no event horizon, its length grows forever at a linear pace. {\it The discovery of this new ``gray phase'' is the main result of the paper.}
\end{itemize}

In the next section, a colliding string wave solution in flat spacetime is described. Section III discusses the case of a string in AdS$_3$ and studies the behavior of the string in the three phases. Section IV describes a simple model that qualitatively explains the numerical results.

\clearpage

\section{Flat spacetime}

As a warmup exercise, let us first construct colliding waves on the string in a flat background geometry.
In (2+1)-dimensional Minkowski space, an explicit solution $X(z,\bar z)$ that describes two colliding waves is given by
\bea
  \nonumber
&  f(z) = a_1 \sin(k_1 z)  \, \theta(z) & \\
  \nonumber
&  g(\bar z) = a_2 \sin(k_2 \bar z) \, \theta(\bar z) &
\eea
\bea
  \nonumber
&  \p \vec X(z,\bar z) = \vec e_1 + \half f(z)^2 \vec e_2 - f(z) \vec e_3 & \\
  \nonumber
&  \bar \p \vec X(z,\bar z) = \vec e_2 + \half g(z)^2 \vec e_1 - g(z) \vec e_3 &
\eea
where $\theta(z)={1\over 1+ e^{-z}}$ is a smooth step function and
\be
  \nonumber
  \vec e_1 = {1 \over \sqrt{2}} (1,1,0);
  \quad   \vec e_2 = {1 \over \sqrt{2}} (1,-1,0);
\quad    \vec e_3 = {1 \over \sqrt{2}} (0,0,1) .
\ee
The spacetime signature is $(-1, 1, 1)$ and the coordinates will be denoted $t, x$ and $y$.
The solution satisfies the equation of motion and the Virasoro constraints,
\be
  \nonumber
  \p \bar\p \vec X = 0 \, , \quad  (\p \vec X)^2 =(\bar \p \vec X)^2 =  0 \, .
\ee

\begin{figure}[h]
\begin{center}
\includegraphics[scale=0.6]{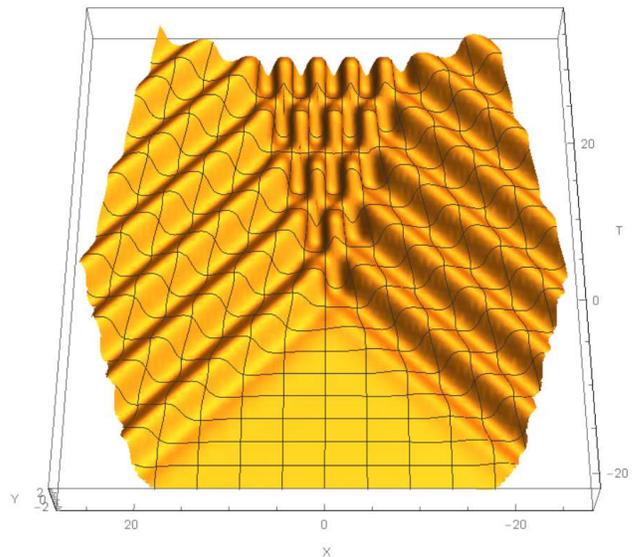}
\caption{\label{fig:flatwaves} Spacetime embedding of the string  at $a=1.5$ and $k=1$. A standing wave region is visible at the top of the figure where the waves collide. The opening angle of this region is smaller than 45$^\circ$.
}
\end{center}
\end{figure}

For simplicity, let us set
\be
  a_1=-a_2=:a \qquad \textrm{and} \qquad k_1=k_2=:k
\ee
Then, $X(z,\bar z)$ can be integrated and expressed in terms of hypergeometric functions. A plot is shown in FIG. \ref{fig:flatwaves}.
Waves initially propagate with the speed of light. When they collide, their effective speed in the background spacetime is slower. This can be seen in FIG. \ref{fig:flatwaves} because the boundary of the standing wave region is not at a 45$^\circ$  angle (which corresponds to the speed of light).

\begin{figure}[h]
\begin{center}
\includegraphics[scale=0.45]{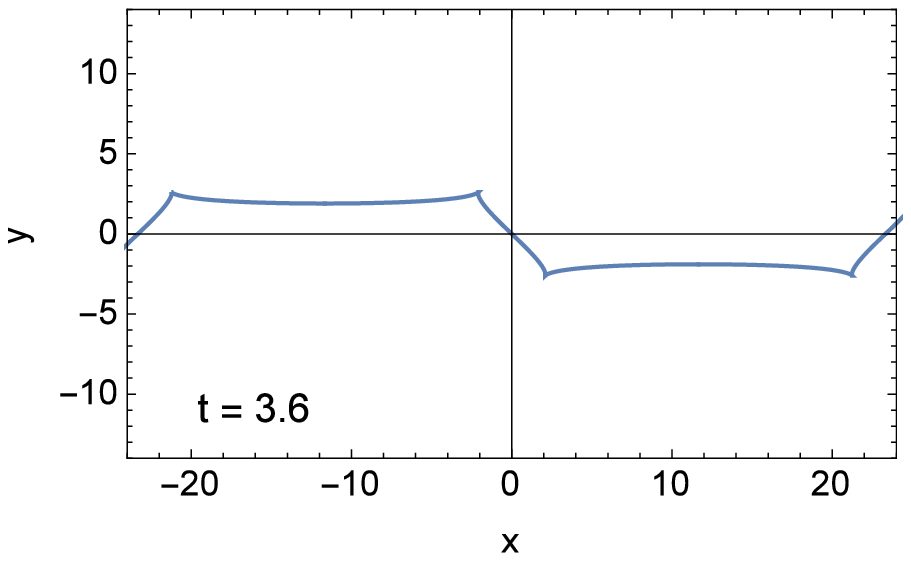}
\includegraphics[scale=0.45]{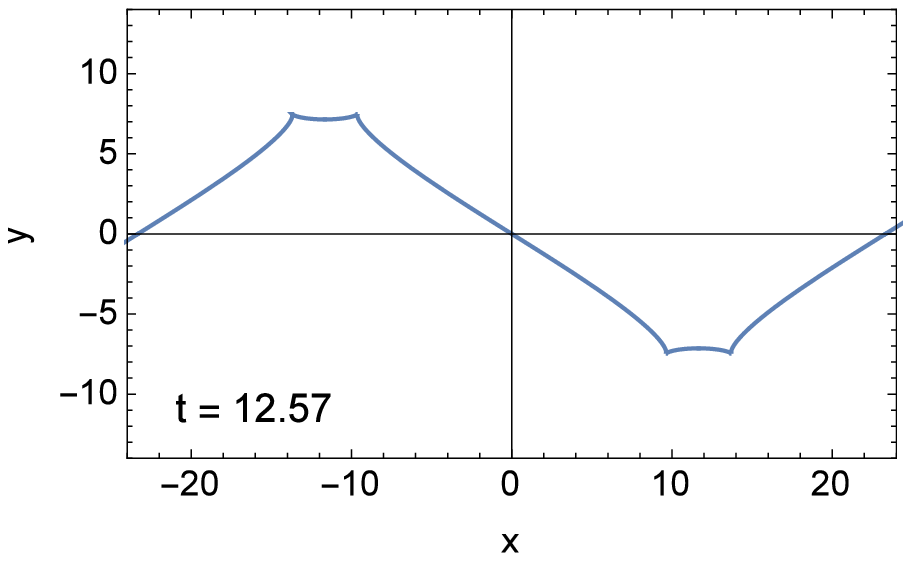} \\
\vskip 0.3cm
\includegraphics[scale=0.45]{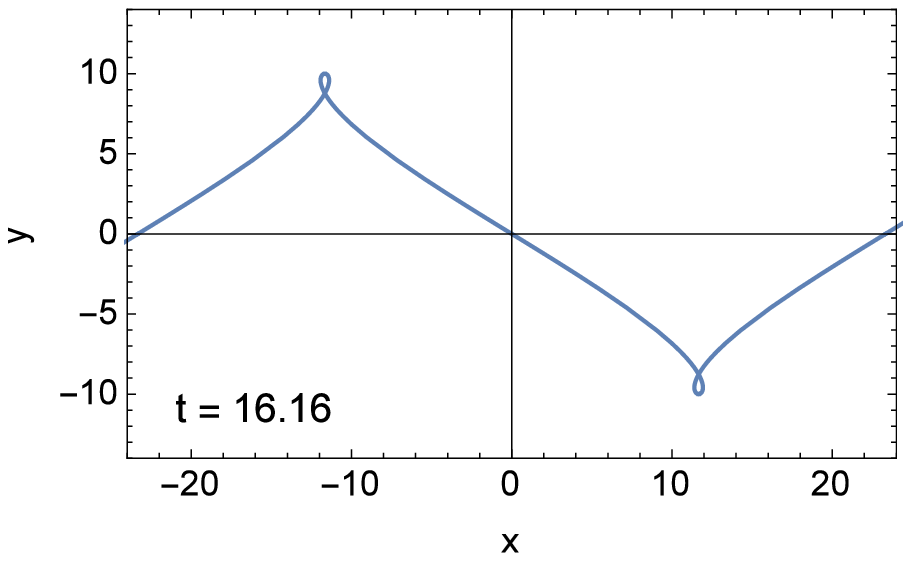}
\includegraphics[scale=0.45]{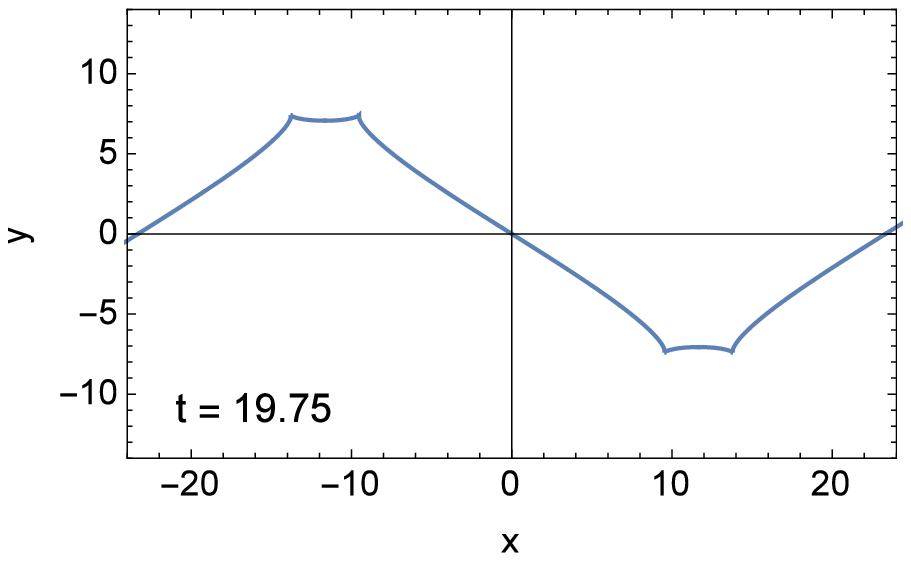} \\
\vskip 0.3cm
\includegraphics[scale=0.45]{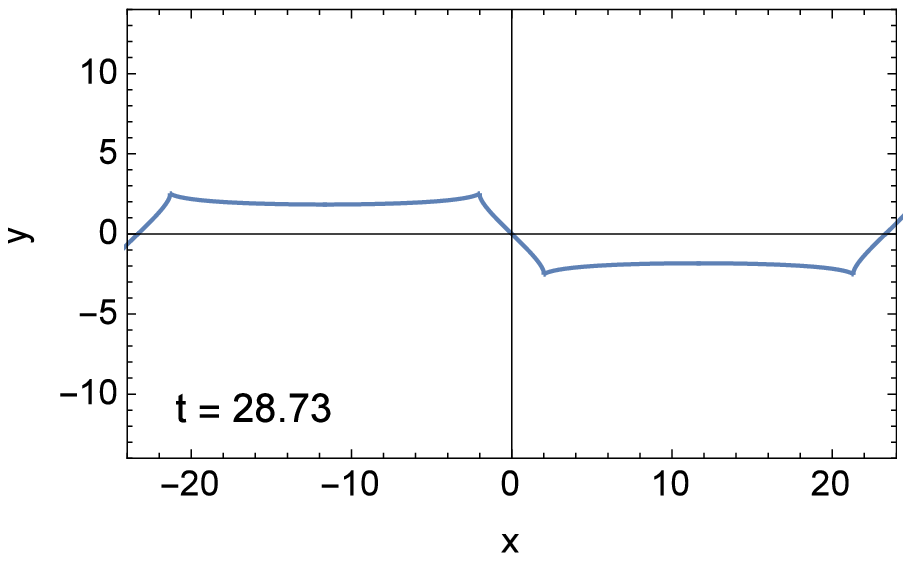}
\includegraphics[scale=0.45]{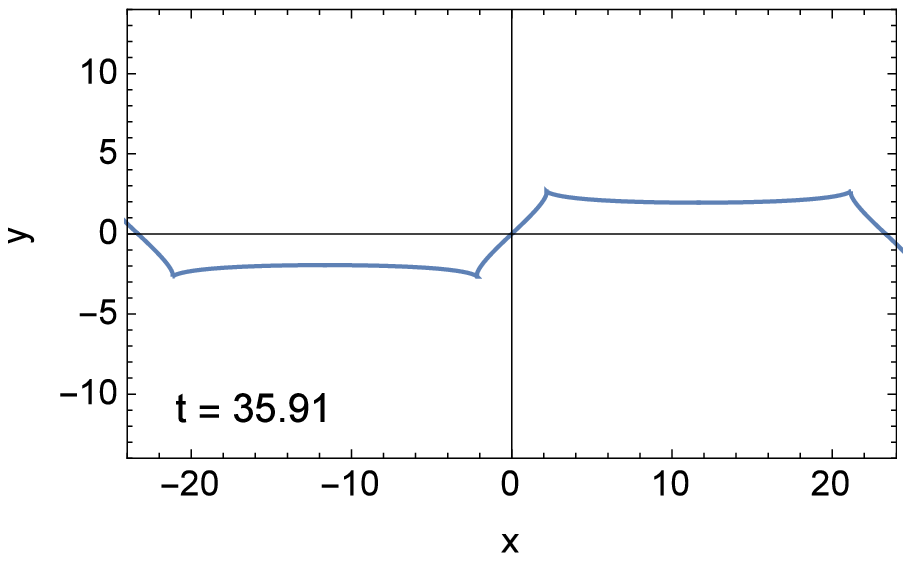}
\caption{\label{fig:behind}
Plots of the string for $a=5.0$ and $k=1$ in the standing wave region (or behind-the-horizon region) at fixed times. Cusps move and collide on the string. The solution is doubly periodic in $t$ and $x$.
}
\end{center}
\end{figure}

If the amplitude of the waves approaches a critical value $a_\textrm{c}=2$, then the standing wave region vanishes. Beyond this value, the opening angle of the standing wave region becomes negative, the string worldsheet folds and the $y(t,x)$ embedding function becomes multi-valued in that region.
The embedding function at late times in in this region can be computed by setting $\theta(z)= 1$. This yields
\bea
  \nonumber
  X^t(\tau, \sigma) = &  {1 \over 4 \sqrt{2}} \le( {(8+2a^2)\tau - a^2 \cos 2\sigma \sin 2\tau } \ri) & \\
   \nonumber
  X^x(\tau, \sigma) =  &  {1 \over 4 \sqrt{2}} \le( {(8-2a^2)\sigma + a^2 \sin 2\sigma \cos 2\tau } \ri)  & \\
  \nonumber
  X^y(\tau, \sigma) =     &   2a \sin\sigma \cos\tau  & .
\eea
where $z=\tau+\sigma$ and $\bar z=\tau-\sigma$. From these implicit formulas, an explicit (but lengthy) expression for $y(t,x)$ can be obtained. A series of plots for $a=5$ at various times is shown in FIG. \ref{fig:behind}. Cusps move on the string and when they collide, the string embedding becomes temporarily smooth (see third plot). The motion is periodic.

\clearpage

\begin{figure}[h]
\begin{center}
\includegraphics[scale=0.45]{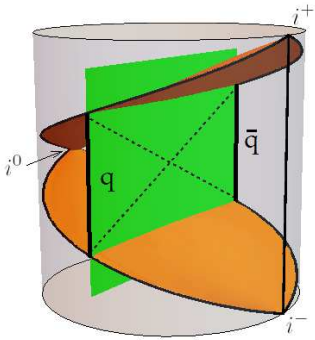}
\caption{\label{fig:globalads}
The basic string solution (in green) in global AdS$_3$ spacetime. The string ends on two quarks ($q$ and $\bar q$) on the boundary. The \poincare patch is bounded by the two orange surfaces.
}
\end{center}
\end{figure}

\section{Anti-de Sitter spacetime}

AdS$_3$ can be embedded into $\RR^{2,2}$  by considering the universal covering space of the surface
\be
  \label{eq:surface}
  \vec Y \cdot \vec Y \equiv -Y_{-1}^2 - Y_0^2 + Y_1^2 + Y_2^2 = -1 .
\ee
The string equations of motion in conformal gauge are
\be
  \label{eq:eoms}
  \p \bar\p \vec Y - (\p \vec Y \cdot \bar\p \vec Y ) \vec Y = 0 .
\ee
which are supplemented by the Virasoro constraints
\be
  \nonumber
  \p \vec Y \cdot \p \vec Y = \bar\p \vec Y \cdot \bar\p \vec Y = 0 .
\ee
The metric on global AdS$_3$ is
\be
  \nonumber
  ds^2 = -{\cosh^2}r  \, dt^2 + dr^2 + {\sinh^2}r \, d\varphi
\ee
where the relationship between $(t, r, \varphi)$ and $Y_i$ is
\bea
  \nonumber
  Y_{-1} &=& \cosh r \cos t \\
  \nonumber
  Y_0 &=& \cosh r \sin t\\
  \nonumber
  Y_1 &=& \sinh r \sin \varphi \\
  \nonumber
  Y_2 &=& \sinh r \cos \varphi
\eea
We are going to plot the string on  \poincare disk time slices. The radial coordinate $\rho \in [0,1)$  of the disk is defined by
\be
  \nonumber
  \rho = \sqrt{\cosh r - 1 \over \cosh r +1}
\ee
and the metric is
\be
  \nonumber
  ds^2 = -\le({1+\rho^2 \over 1-\rho^2} \ri)^2  \, dt^2 + {4  \over (1-\rho^2)^2} \le(  d\rho^2 + \rho^2 \, d\varphi \ri)
%  ds^2 = -{\cosh^2}r  \, d\tau^2 + 4{ d\rho^2 + \rho^2 \, d\varphi  \over (1-\rho^2)^2}
\ee

\begin{figure}[h]
\begin{center}
\includegraphics[scale=0.35]{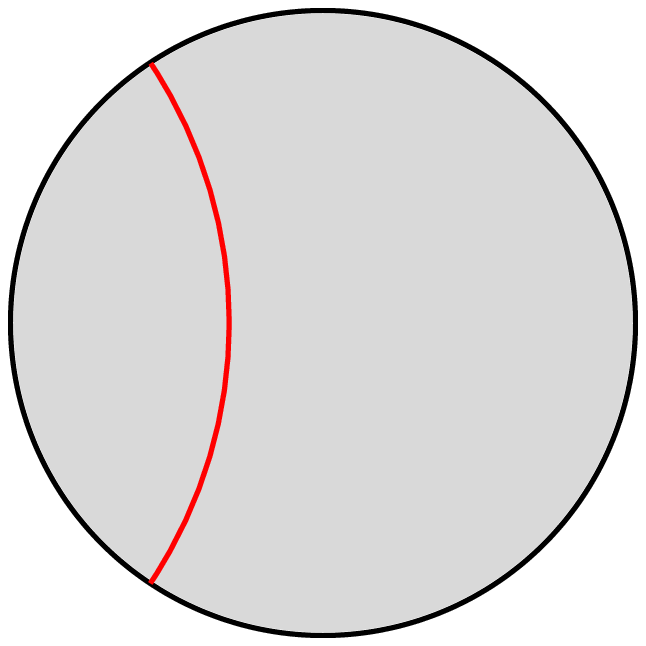}
\includegraphics[scale=0.35]{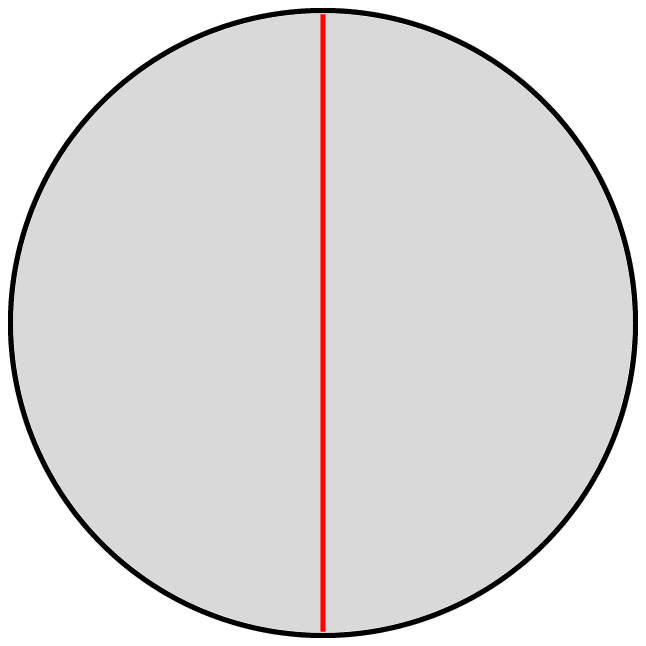}
\includegraphics[scale=0.35]{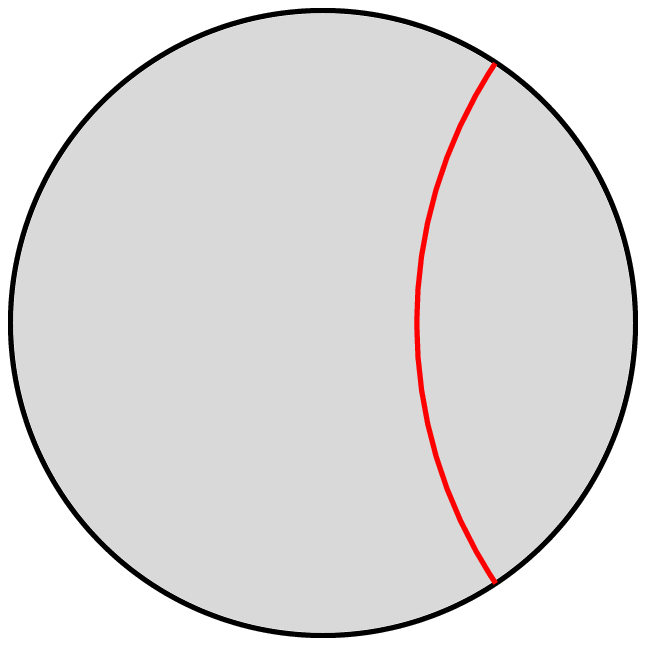}
\caption{\label{fig:rock}
Boosted string with a constant normal vector in global AdS. The three figures are \poincare disk timeslices at $t = 0, {\pi \over 2}$, and $ {\pi}$. The string (shown in red) oscillates in the left-right direction with a period of $2\pi$. The shape is a circular arc, perpendicular to the boundary circle (on the Euclidean plane).
}
\end{center}
\end{figure}

A normal vector to the string can be defined as
\be
  \nonumber
  N_a = { \epsilon_{abcd} Y^b \p Y^c \bar\p Y^d \over \p \vec Y \cdot \bar\p \vec Y}
\ee
It satisfies $\vec N \cdot \vec Y =\vec N \cdot \p \vec Y = \vec N \cdot \bar\p \vec Y  = 0 $ and $\vec N \cdot \vec N = 1$.
A simple solution to the equation of motion is obtained by setting $N$ to be a constant vector. It is the AdS$_3$ analog of an infinitely long straight string in flat spacetime. The string embedding corresponding to  $\vec N(t) = (0,0,0,1)$ is shown in FIG. \ref{fig:globalads}.

By applying an appropriate  transformation from the $SO(2,2)$ isometry group of AdS$_3$, the string can be boosted and/or rotated. A boosted string will oscillate in $AdS$ with a period of $2\pi$. This is seen in FIG. \ref{fig:rock}. At a given time, the string embedding on the 2d plane parametrized by $\rho$ and $\varphi$ is a circle arc that is perpendicular at both endpoints to the boundary circle.
The boundary circle has a unit radius and it is centered at the origin. The center in polar coordinates ($\rho_0$ and $\varphi_0$) and radius ($r_0$) of the circle arc can be computed in terms of its normal vector $\vec N = (n_1, n_2, n_3, n_4)$ to be
\bea
  \nonumber
  & \tan\, \varphi_0 =  {n_3 \over n_4} &  \\
  \nonumber
   & \rho_0 = {1+R^2 \over  \tilde n_1} \sqrt{{n_1^2 + n_2^2}} &  \\
  \nonumber
   & r_0  = \sqrt{4 R^2 + (1+R^2)^2 {\tilde n_2 \over \tilde n_1} } &
\eea
where
\be
  R = {1 \over { \sqrt{n_1^2 + n_2^2} +\sqrt{n_3^2 + n_4^2} } }
\ee
and the two dimensional vector $(\tilde n_1, \tilde n_2)$ rotates with global AdS time,
\be
\nonumber
 \binom{\tilde n_1}{\tilde n_2} = \left(
\begin{array}{cc}
\cos t &  -\sin t \\
\sin t & \cos t
\end{array}
\right) \binom{n_1}{n_2}
\ee

\begin{figure}[h]
\begin{center}
\includegraphics[scale=0.35]{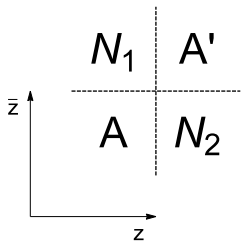}
\caption{\label{fig:latticeexpl} Four patches on the worldsheet. The dashed lines  in the middle are lightlike worldlines of two colliding cusps. The normal vectors are  labeled $\vec A$, $\vec A'$, $\vec N_1$, and $\vec N_2$. The collision formula computes any one of these vectors from  the other three.
}
\end{center}
\end{figure}

More complex strings can be glued from such ``straight'' pieces. At the gluing points the string will contain cusps that can collide.
In AdS$_3$, the situation on the worldsheet is shown in FIG. \ref{fig:latticeexpl}. Time grows in the up/right direction. The two dashed lines are the worldlines of the cusps. Before the collision, the string consists of three pieces that are characterized by three normal vectors: $\vec N_1$, $\vec A$ and $\vec N_2$. We require $ \vec A \cdot \vec N_1 = \vec A \cdot \vec N_2 = 1$ so that the cusps  move with the speed of light.
After the collision, $\vec A$ changes to $\vec A'$ given by the collision formula
\be
  \label{eq:reflection}
  \vec A' = -\vec A +   4 {\vec N_1 + \vec N_2 \over (\vec N_1 + \vec N_2)^2 }
\ee
Further cusp collisions can be computed by repeated applications of the formula.

\subsection{The setup}

\begin{figure}[h]
\begin{center}
\includegraphics[scale=0.38]{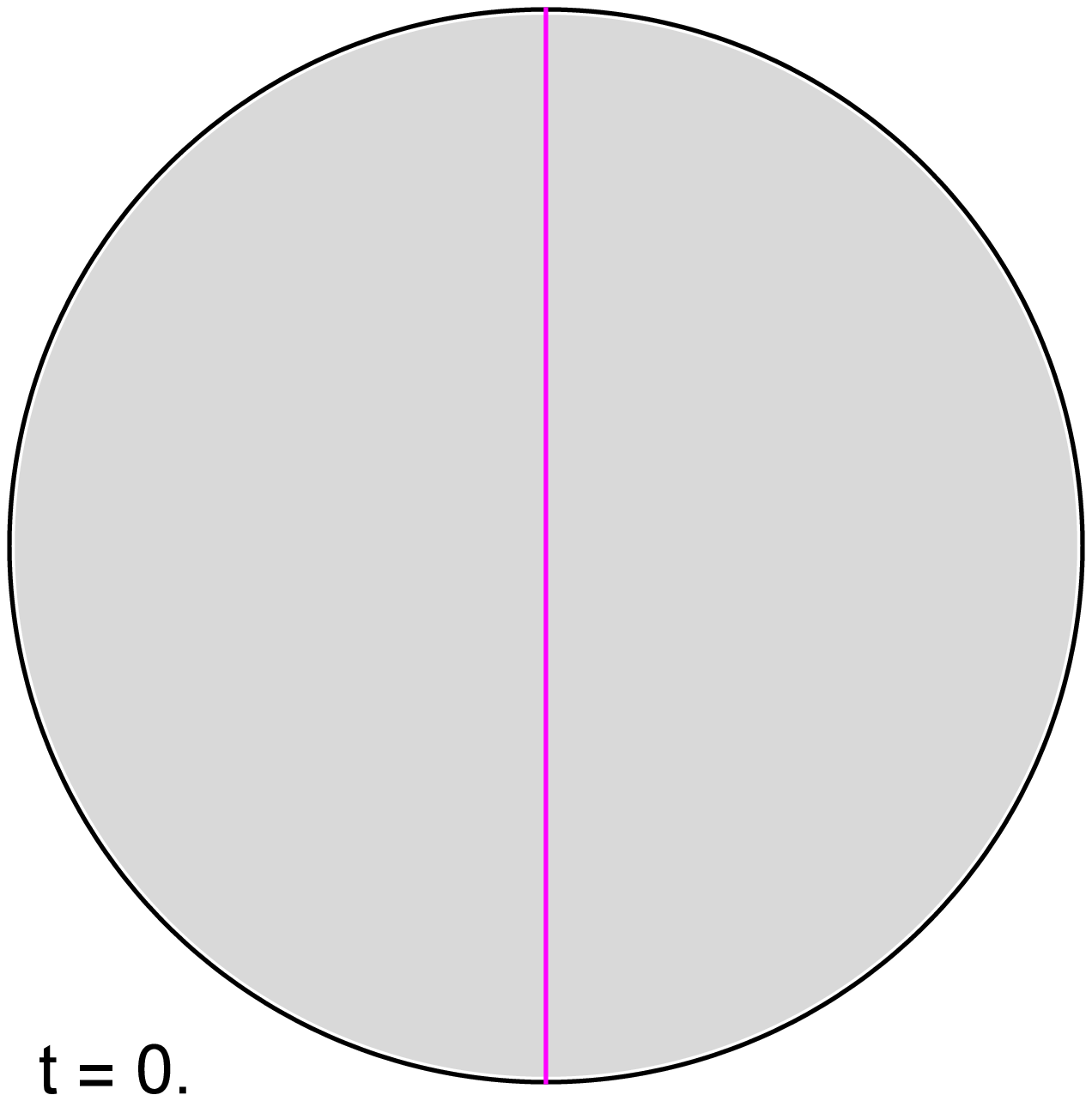}~\includegraphics[scale=0.38]{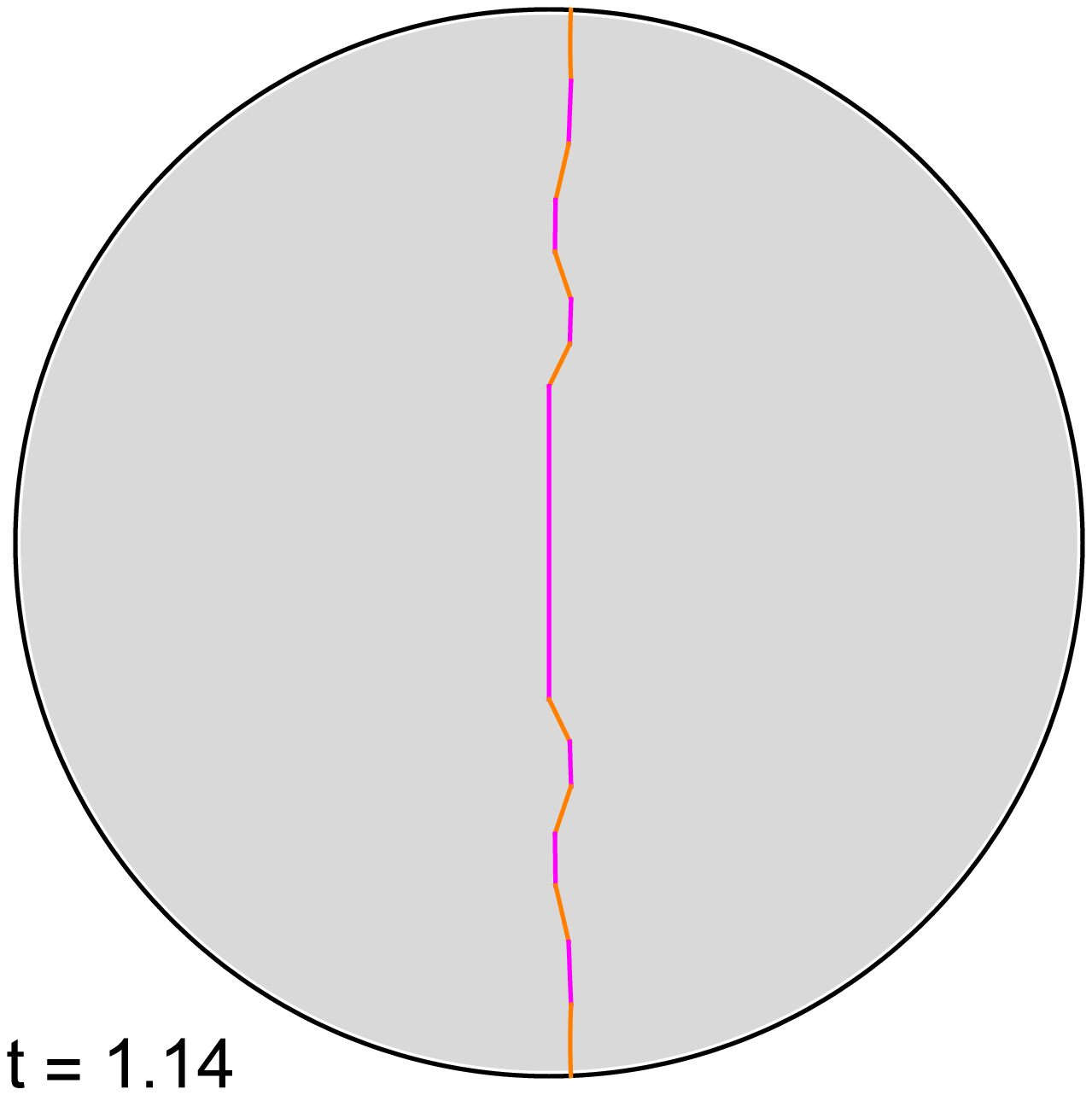}~\includegraphics[scale=0.38]{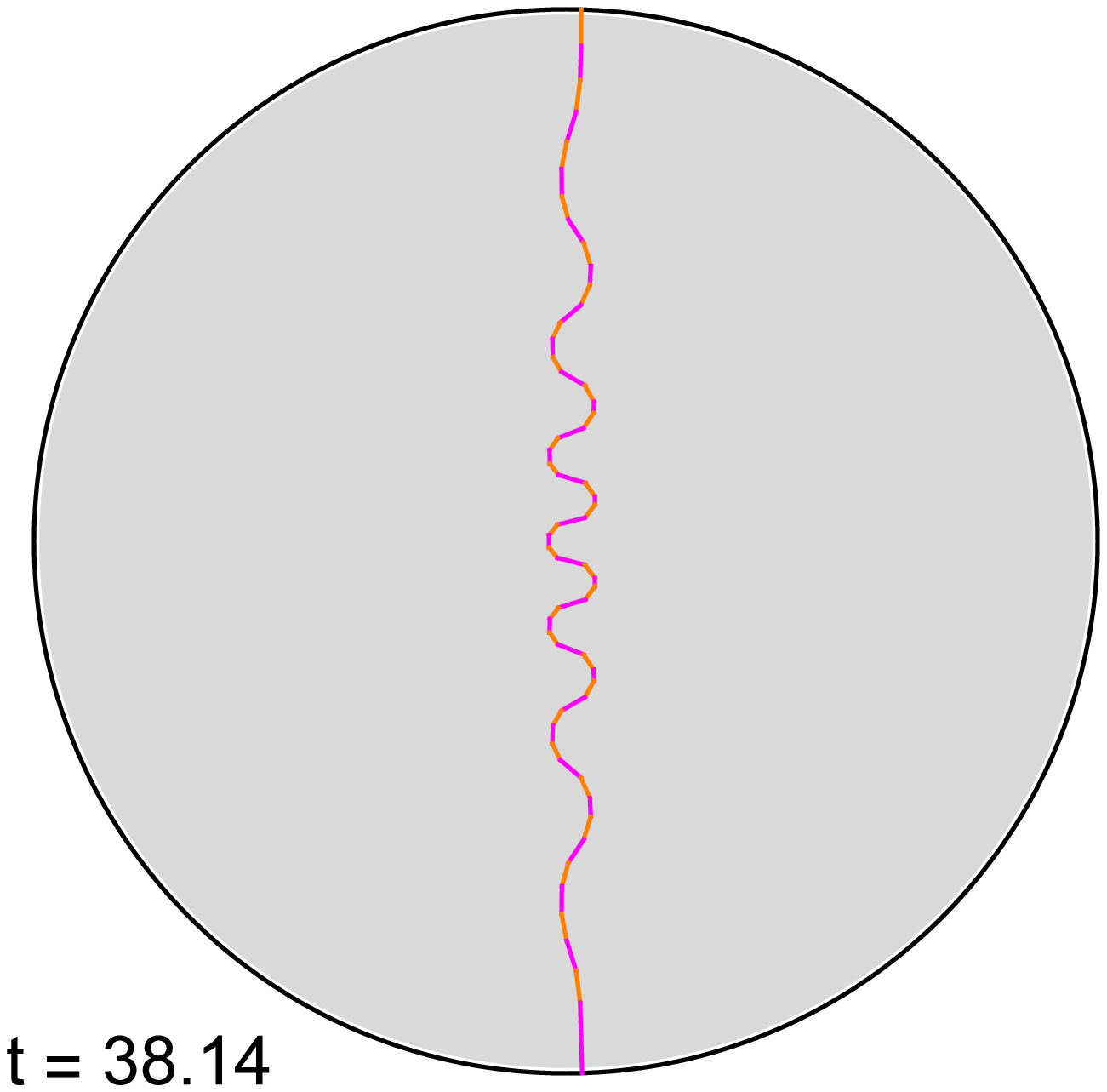}
\caption{\label{fig:adsconstant}
Shockwave collision at small kick strengths ($\nu=0.6$). The color of string segments alternates for enhanced visibility. After some time, a standing wave forms and the motion is visibly periodic. The animation is available at \cite{website}.}
\end{center}
\end{figure}

Let us now stretch a string between two antipodal points on the boundary. The first picture in FIG. \ref{fig:adsconstant} shows this embedding. The corresponding normal vector is constant: $\vec N_0 = (0,0,0,1)$. Waves can be sent in by perturbing the two endpoints (see  \cite{Ishii:2015wua} for a related calculation on the \poincare patch). The waves collide at $t={\pi \over 2}$ and scatter from each other.

From a technical standpoint, the simplest waves consist of cusps and boosted straight string segments in between. For instance, one could consider a left-right oscillating endpoint that would generate a triangle wave on the string. The quark on the boundary suffers equal kicks in alternating directions.

The quarks will be kicked as follows.
Let us consider the string segment that ends on the quark. Let $\varphi_i(t)$ parametrize the quark position just before the i$^\textrm{th}$ kick. This is an angle on the boundary circle of the \poincare disk. By kicking the quark we really mean that we let its acceleration jump by a finite value. This creates a cusp at the boundary and it starts moving toward the interior. Behind the cusp a new string piece is created whose endpoint angle will be denoted by $\varphi_{i+1}$. At the time of the kick, this new segment satisfies
\be
   \varphi_{i+1}(t_i) = \varphi_i(t_i) \qquad \dot\varphi_{i+1}(t_i) = \dot\varphi_i(t_i)
\ee

\begin{figure}[h]
\begin{center}
\includegraphics[scale=0.5]{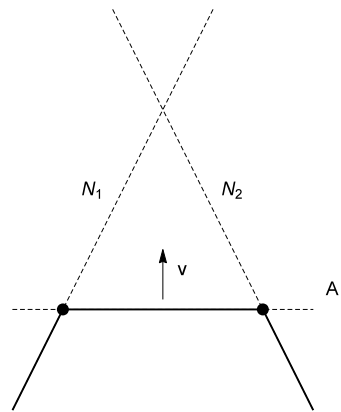}
\caption{\label{fig:expl} Collision of cusps on the string (thick line). In an appropriate frame the lines $N_1$ and $N_2$ are static. The collision inverts the velocity of $A$: $\vec v \to -\vec v$.
}
\end{center}
\end{figure}
and
\be
  \nonumber
  \le.{dp_{i+1} \over dt}\ri|_{t=t_i} =  \le.{dp_{i} \over dt}\ri|_{t=t_i} + \nu_i
\ee
where we have defined $p_i = {\dot\varphi_i\over \sqrt{1-\dot\varphi_i^2}}$ and  $\nu_i$ is the strength of the i$^\textrm{th}$ kick. This leads to particularly simple formulas.
Finally, whenever an outgoing cusp reaches the AdS boundary, the outermost string segment is removed from the system. This allows cusps to exit the string.

In the remainder of this paper, we will consider square waves: a kick of the quark, then two kicks in the opposite direction, then two kicks in the original direction and so on. This way the entire string will stay approximately at the same position without boosting the initial string. The kicks start at $t=0$ and continue at equal time spacing with respect to global AdS time. In the following, this time spacing will be set to a fixed value $\Delta t = {1\over 7}$ and only the kick strength will be varied.

The simulations have been done using Wolfram's Mathematica \cite{ram2014}.
The algorithm is based on the author's public code in \cite{Vegh:2015ska} with a few modifications:
\begin{itemize}
\item
The code has been extended to handle string segments and cusps in global anti-de Sitter spacetime.
\item
Cusps are added according to the prescription above.
\item
Cusps are removed when they reach the boundary.
\item
Due to the number of cusps, a projection has to be performed on the $ \vec N_i \cdot \vec N_j = 1$ subspace where $i$ and $j$ denote two neighboring string segments. The projection is necessary, because real numbers are stored digitally and the rounding errors would otherwise grow exponentially with the number of collisions.
\item
Finally, the string is plotted on  \poincare disk timeslices.
\end{itemize}

\clearpage

\bwt
\onecolumngrid

\begin{figure}[h]
\begin{center}
\includegraphics[scale=0.42]{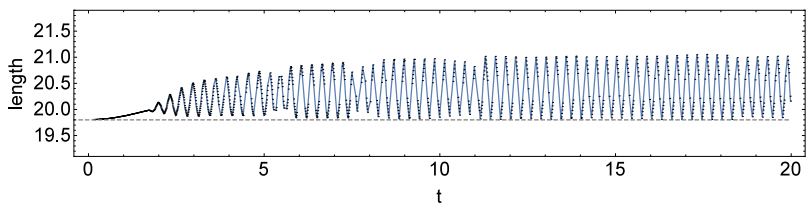}
\caption{\label{fig:length1}
String length in the transparent phase. The length fluctuates around a constant value as the standing wave oscillates. The initial length of the string (between the UV cutoffs) is shown by a dashed line.
}
\end{center}
\end{figure}

\twocolumngrid

\ewt

\subsection{Transparent phase}

For weak kicks, an example is shown in FIG. \ref{fig:adsconstant}. Three time slices are plotted: (i) the initially static string, (ii) after the start, waves start to propagate, and finally (iii) after some time a standing wave forms\footnote{This is similar to the linearized case where the resulting wave is simply the sum of two waves traveling in opposite direction.}.
Cusps enter the system as the quarks are kicked, they travel through the middle part and they exit at the other endpoint. Thus, this phase is termed {\it transparent}. There is an (approximately) constant number of cusps on the string at any given time.

\begin{figure}[h]
\begin{center}
\includegraphics[scale=0.7]{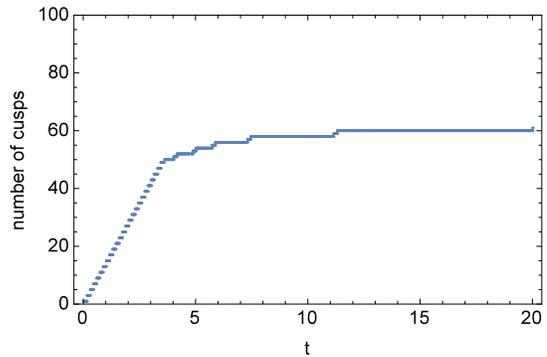}
\caption{\label{fig:number1}
Number of cusps on the string in the transparent phase. At late times, in the steady state, it approaches a constant.
}
\end{center}
\end{figure}

The string length is shown in FIG. \ref{fig:length1}. Here the length is defined on a timeslice as the sum of lengths of all individual string segments\footnote{Generically, this is not the minimal distance between the two endpoints.}. The geodesic length  of a single segment is given by
\be
  \nonumber
  s(\vec X, \vec Y) = \log\le[ -  \vec X \cdot \vec Y + \sqrt{(\vec X \cdot \vec Y)^2-1} \ri]
%  e^{s(\vec X, \vec Y)} =   -  \vec X \cdot \vec Y + \sqrt{(\vec X \cdot \vec Y)^2-1}
\ee
where $\vec X$ and $\vec Y$ denote the two endpoints of the segment on the hyperbola (\ref{eq:surface}) in the embedding space. In order to have a finite result, the string is cut off in the ultraviolet. In this paper the cutoff is set to $\rho_\textrm{cutoff} = 1-10^{-4}$. Then, the initial length is $s_0 = 4 \, \textrm{artanh} \, \rho_\textrm{cutoff} \approx 19.8$ which is indicated by a dashed line in the figure.

The number of cusps as a function of global AdS time is shown in FIG. \ref{fig:number1}. In the steady state, the cusp number saturates (in this example around 60 cusps). This means that cusps enter and leave the system at the same rate.

\clearpage

\bwt
\onecolumngrid
\begin{figure}[h]
\begin{center}
\includegraphics[scale=0.38]{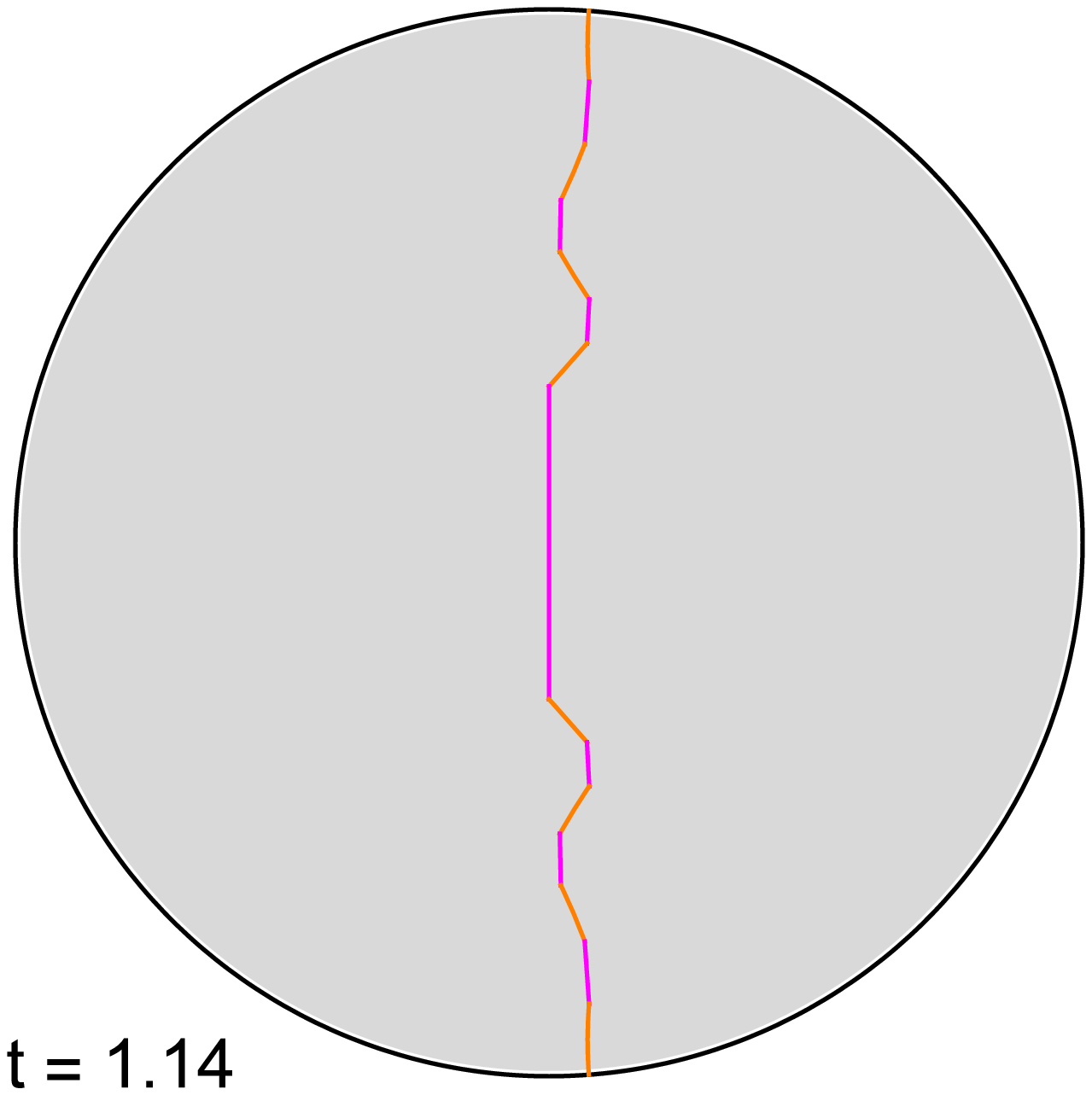}~\includegraphics[scale=0.38]{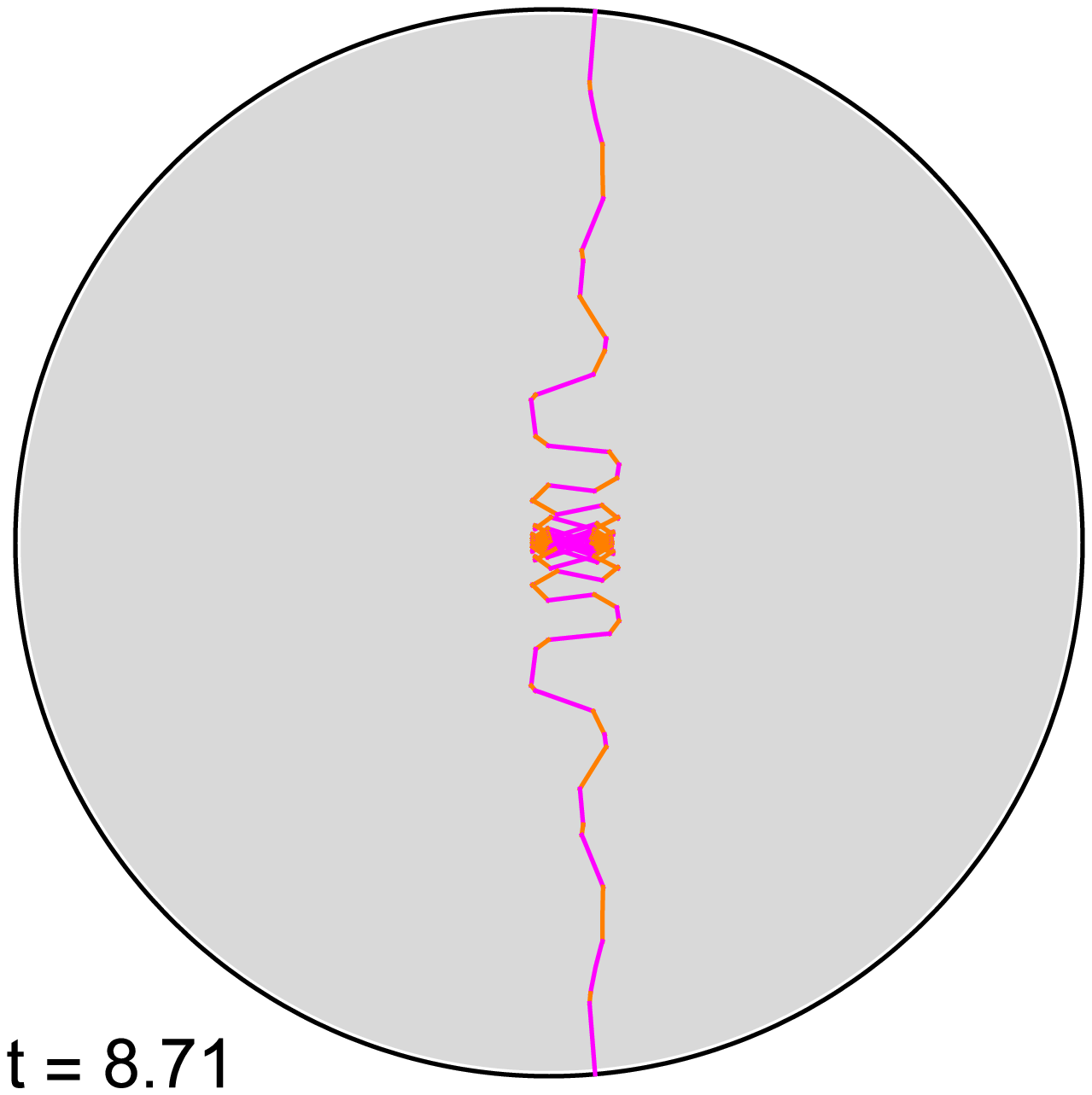}~\includegraphics[scale=0.38]{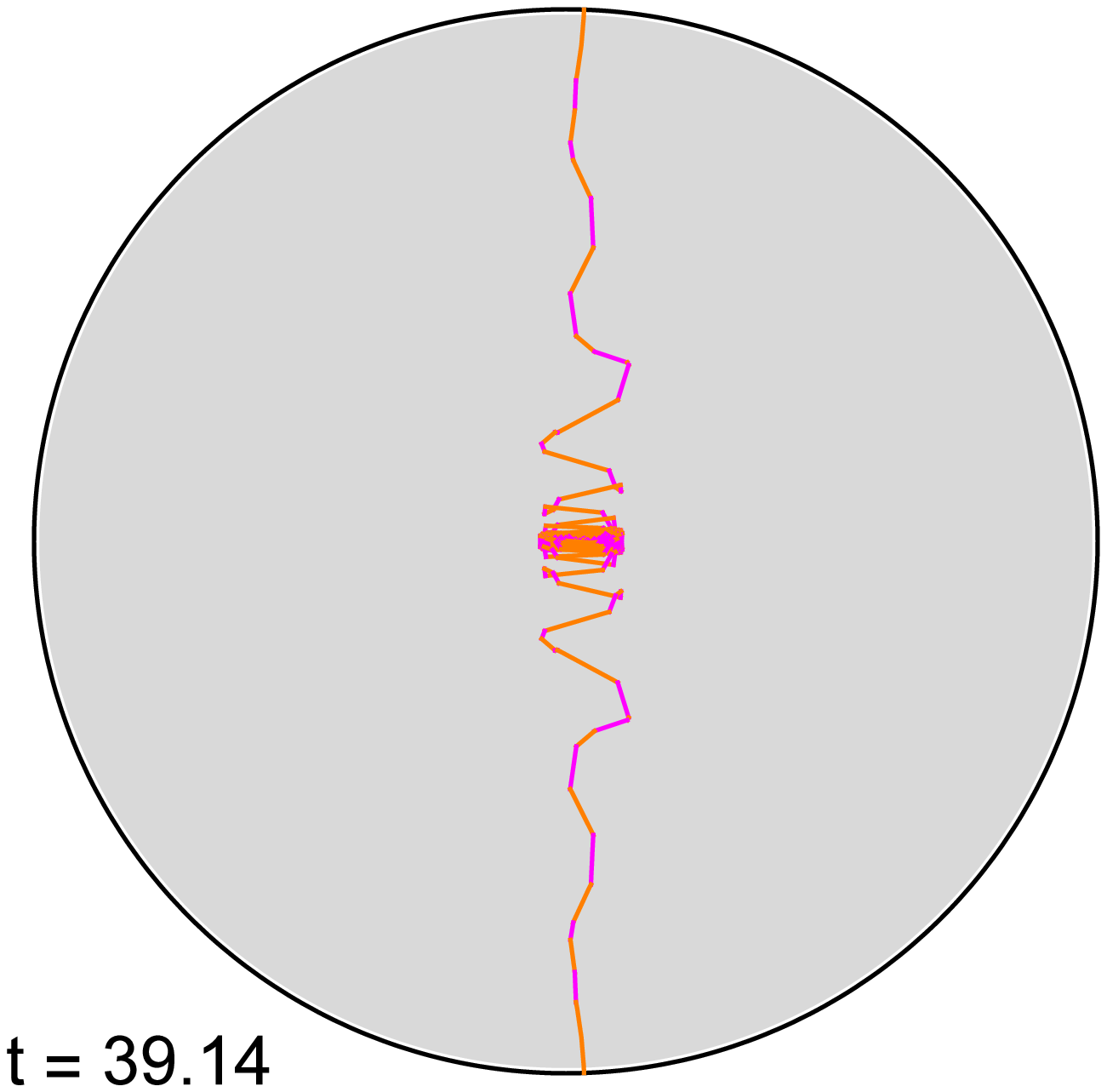}
\caption{\label{fig:adslinear}
Shockwave collision at intermediate kick strengths ($\nu=1.1$). Cusps pass through the string, but the motion is not periodic. The string length increases linearly over time and cusps accumulate in the center. The animation is available at \cite{website}.
}
\end{center}
\end{figure}

\begin{figure}[h]
\begin{center}
\includegraphics[scale=0.45]{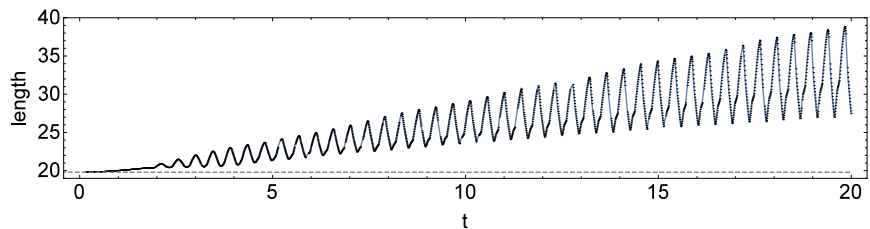}
\caption{\label{fig:length2}
String length in the gray phase. The length fluctuates similarly to the transparent case. However, the average length grows over time.
}
\end{center}
\end{figure}
\twocolumngrid

\ewt

\begin{figure}[h]
\begin{center}
\includegraphics[scale=0.7]{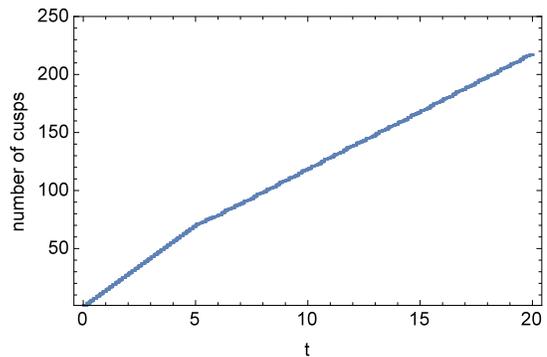}
\caption{\label{fig:number2}
Number of cusps on the string in the gray phase. The number of cusps grows linearly with time until the first cusps exit on the other end (here around $t \approx 5$). Then, the string continues growing, albeit at a somewhat smaller rate. The number of cusps grows linearly after a short transient.
}
\end{center}
\end{figure}

\subsection{Gray phase}

If one increases the strength of kicks, the system enters a new phase. An example is shown in FIG. \ref{fig:adslinear}. Three time slices are plotted. The string keeps folding in the center of AdS and becomes longer and longer. Cusps do cross this region though and they exit at the other end. This region will be called the {\it gray phase}.

The string length grows, see FIG. \ref{fig:length2}. The number of cusps grows linearly for a while, then as cusps start exiting at the other end, the growth rate decreases. This is shown in FIG. \ref{fig:number2}. The numerical calculation shows that the cusp number in the steady state grows linearly over time.

The transparent/gray transition happens around $\nu\approx 0.71$.

\clearpage

\bwt
\onecolumngrid

\begin{figure}[h]
\begin{center}
\includegraphics[scale=0.38]{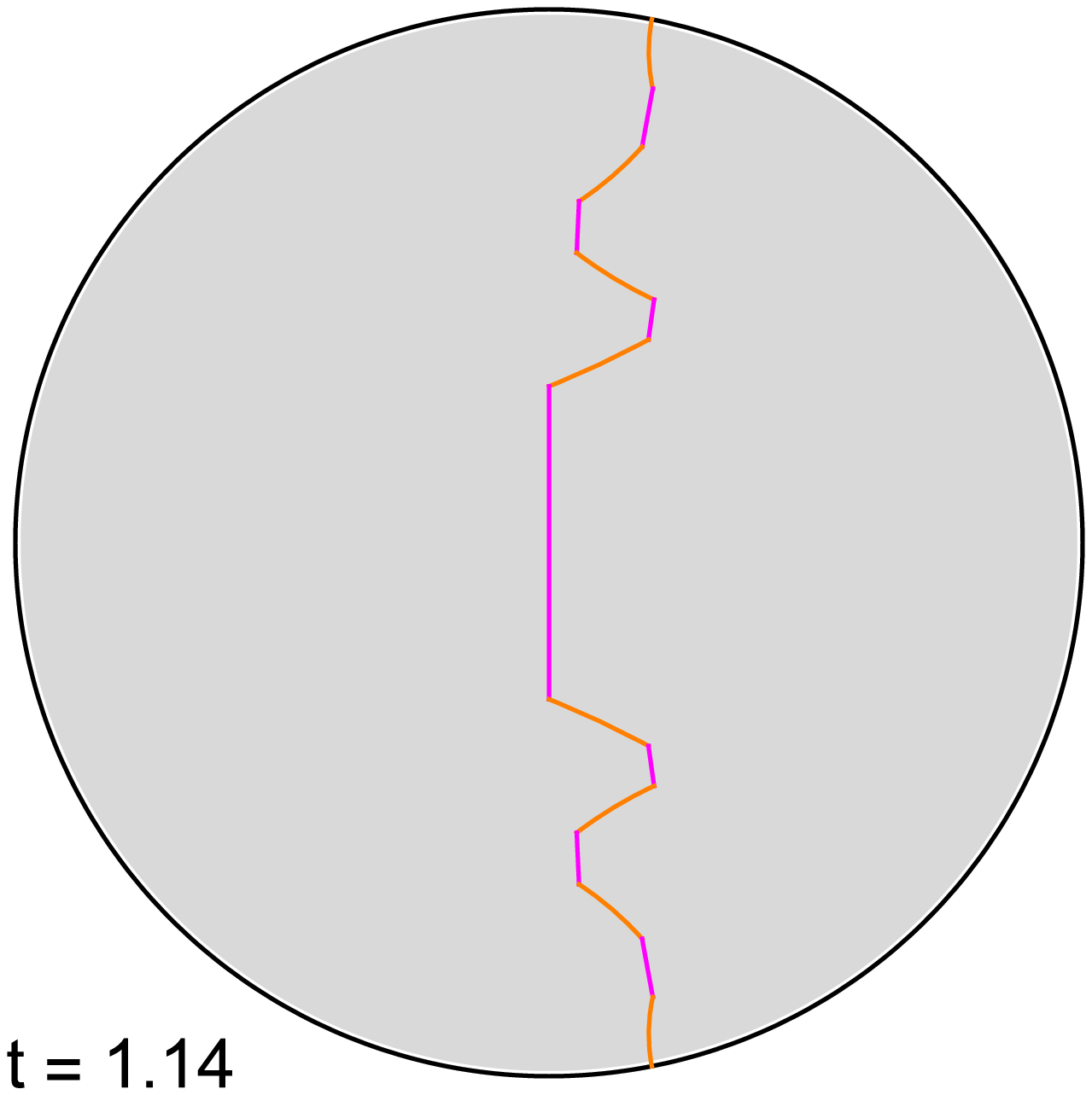}~\includegraphics[scale=0.38]{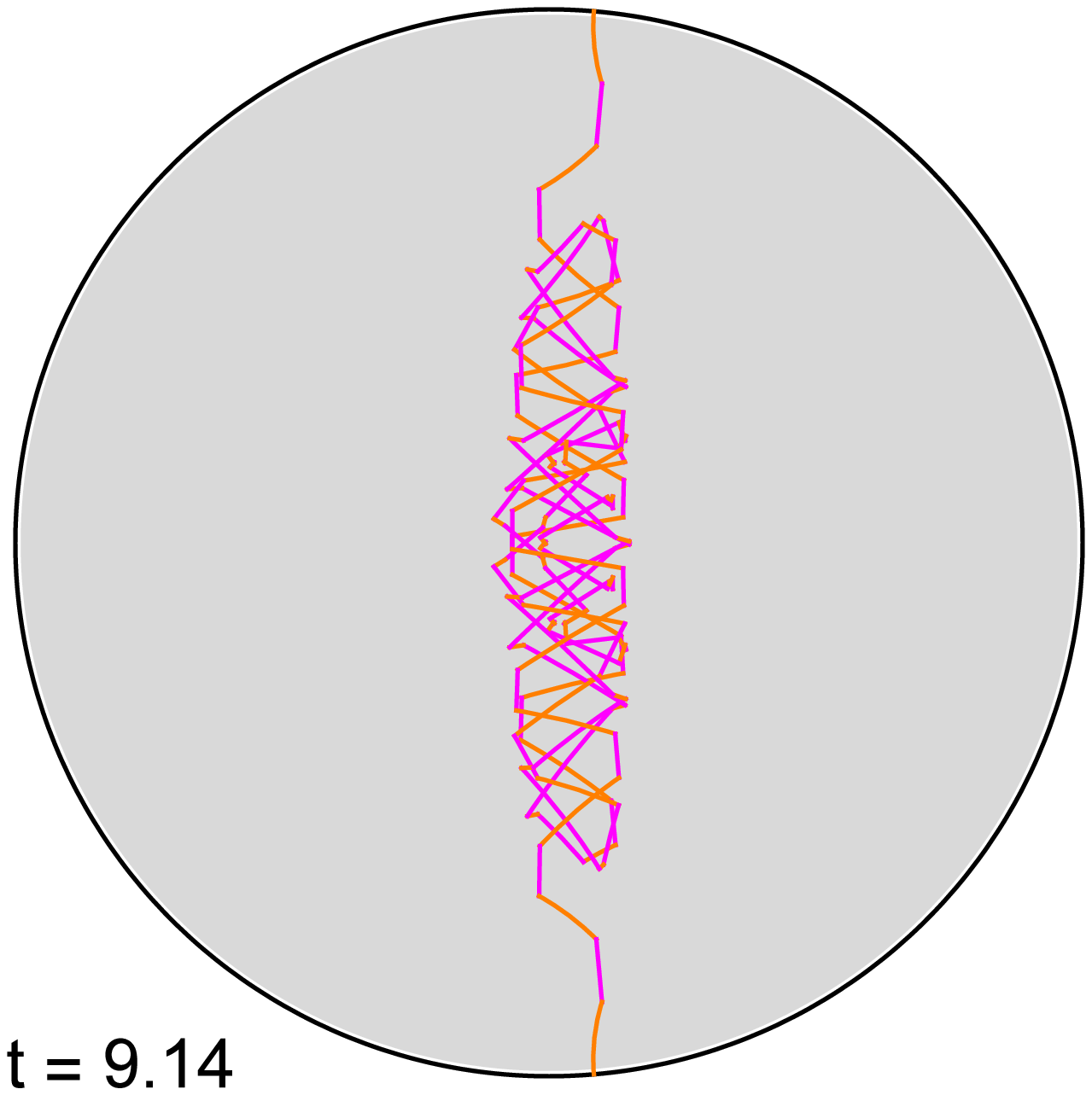}~\includegraphics[scale=0.38]{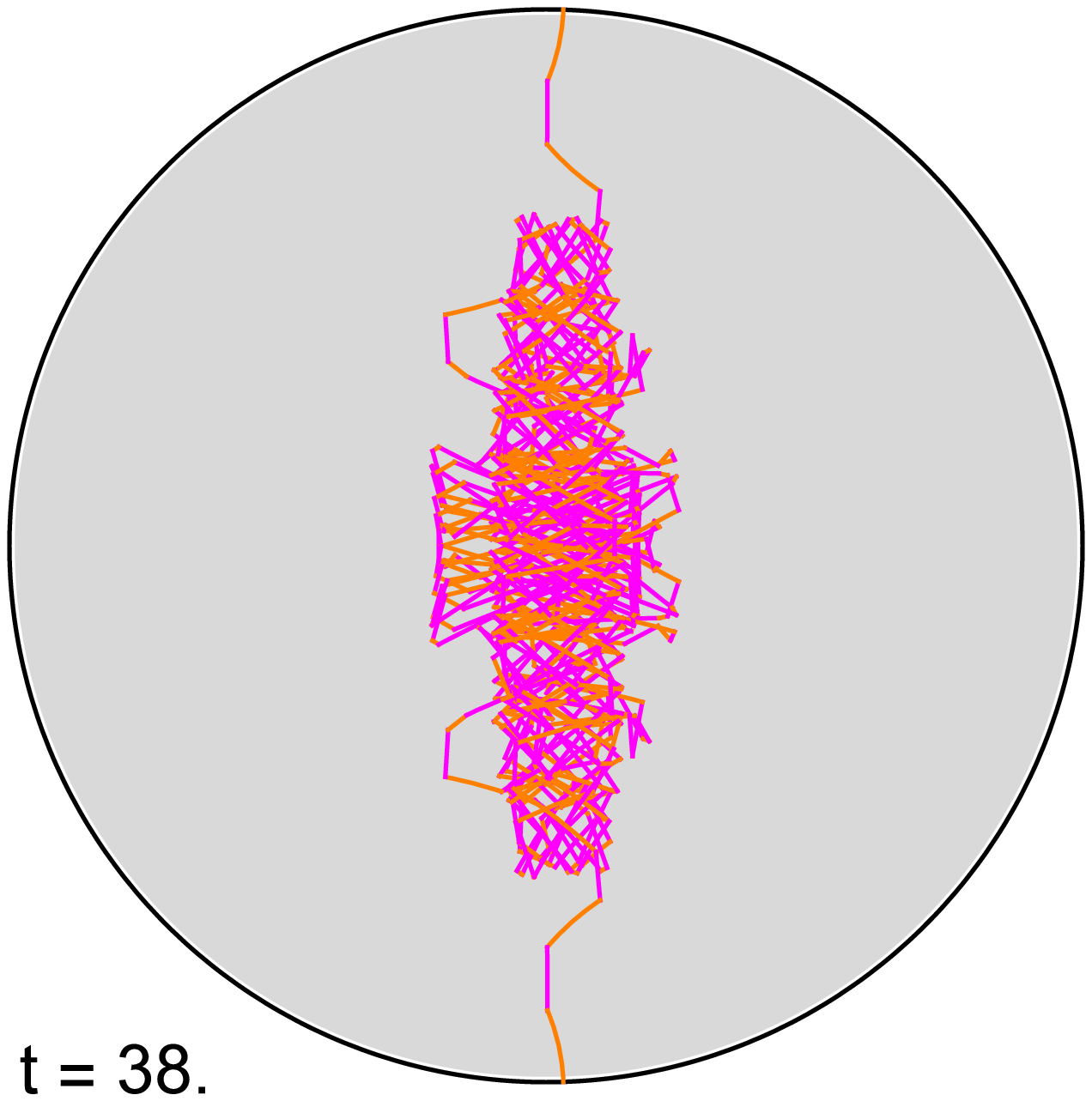}
\caption{\label{fig:adsbh}
Shockwave collision in the black phase at large kick strengths (here $\nu=3$). There are two worldsheet horizons and at late times the string fills in a region of the bulk.  The animation is available at \cite{website}.
}
\end{center}
\end{figure}

\begin{figure}[h]
\begin{center}
\includegraphics[scale=0.45]{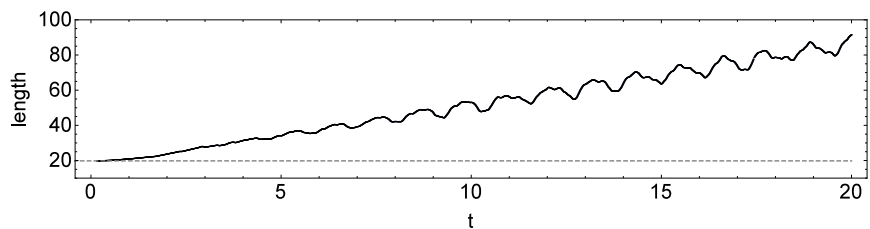}
\caption{\label{fig:length3}
String length in the black phase.
}
\end{center}
\end{figure}

\twocolumngrid

\ewt

\subsection{Black phase}

If the kick strength is further increased, eventually an event horizon forms on the worldsheet. The cusps get stuck in the bulk and the string keeps folding ad infinitum.
An example is shown in FIG. \ref{fig:adsbh}. The string length is plotted in FIG. \ref{fig:length3}.

Even though we have not provided a mathematical proof for the existence of the horizon, long simulations (up to $t \approx 150$) show strong evidence for this. Horizons are seen to form whenever the collision of the first two cusps results in a folded string. Then, further cusps continue to fold the string which simply gets longer at a rate faster than the speed of light. Thus, cusps cannot reach the other side. In our example of the ``square wave'', the horizon is determined to appear around $\nu\approx 1.41$.

\subsection{Phase transitions}

\begin{figure}[h]
\begin{center}
\includegraphics[scale=0.6]{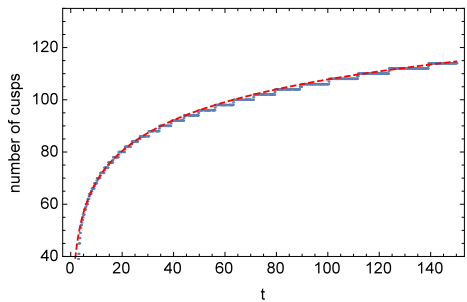}
\caption{\label{fig:number3}
Time evolution of the number of cusps on the string in the transparent/gray phase transition point. The data points are plotted in blue. The red curve is a simple logarithmic fit. The function starts out linearly before the first cusps exit on the other end, but this is not visible in the figure.
}
\end{center}
\end{figure}

\begin{figure}[h]
\begin{center}
\includegraphics[scale=0.55]{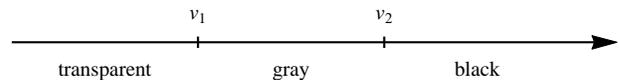}
\caption{\label{fig:param}
A rudimentary phase diagram where only the strength of kicks $\nu$ is varied. For $\nu < \nu_1$ the system is transparent. At intermediate $\nu_1 < \nu < \nu_2$ the string keeps growing over time (gray phase). At large $\nu>\nu_2$ the worldsheet horizon prevents cusps from reaching the other side (black phase).
}
\end{center}
\end{figure}

A rudimentary phase diagram is shown in FIG. \ref{fig:param}.
What is the nature of the phase transition between the transparent and gray regions?
FIG. \ref{fig:number1} and \ref{fig:number2} show the change in cusp numbers in the two cases.
Both figures show a linear increase in the beginning, then the cusp number either saturates or continues linearly in the transparent and gray cases, respectively. Let us concentrate on the gray phase. The elbow in FIG. \ref{fig:number2} indicates the moment when the first cusps exit on the other end of the string. Now the parameters can be tuned so that the system  moves toward either the black or the transparent phase. Thus, there are two options:
\begin{itemize}
\item
As the parameters approach the black (horizon) region, the elbow in FIG. \ref{fig:number2} moves to the right and goes off to infinity. This signals the formation of the worldsheet horizon.
\item
Right after the elbow, there is a transient region where the growth is not linear yet. If the system moves closer to the transparent phase, then this transient region lasts longer and eventually diverges. The resulting plot is shown in blue in FIG. \ref{fig:number3}. The dashed red line is a logarithmic fit (power law functions were not a good fit).
\end{itemize}

%\vskip -0.35cm

%\clearpage

\section{A simple model}

The scattering of many cusps is a complex process, but the steady state results are fairly robust. There are three distinct phases with different characteristics. What is the simplest model that incorporates these features?

Here we discuss a simple model that reproduces the three phases\footnote{I thank Douglas Stanford for suggesting this model to me.}. Let us consider cusps of equal $\nu$ strength that enter the string at every second on both side. The model for the ``Shapiro time delay'' through the cusp is simply that it increases the string length by $\nu$. In the following,  the initial ($t\leq 0$) string length will be neglected. If at a given time the string contains $m$ cusps, its length will be taken to be $m \nu$. Let $n(t)$ denote the number of cusps that have already exited by the time $t$. Let us now consider a right-moving cusp that enters the string from the left at $t=t_\textrm{in}$ and exits on the right at $t_\textrm{out}$. As it travels through the string, the cusp has to cross at most $t_\textrm{out}$ cusps that entered from the right since $t=0$. However, some of these left-moving cusps have already exited before $t_\textrm{in}$ and their number must be subtracted. We then have the equation
\be
  t_\textrm{out} - t_\textrm{in} = \nu \, t_\textrm{out} - \nu \, n(t_\textrm{in})
\ee
From this one gets
\be
  t_\textrm{out}(t_\textrm{in}) = {t_\textrm{in}- \nu \, n(t_\textrm{in}) \over 1-\nu}
\ee
This formula computes the time of ``response'' in terms of the time of the kick.
Using this expression, one can compute $n(t)$ by integrating
\be
  n'(t) = \sum_{t_\textrm{in}=0,1,2,\ldots} \delta\le[t- t_\textrm{out}(t_\textrm{in}) \ri]
\ee
This can be done numerically. (A simple Mathematica code that does the job is attached to the paper.)

Finally, the cusp number $N(t)$  is simply the number of cusps that have entered the system minus the number of those that have left already,
\be
  N(t) = 2(t-n(t))
\ee
Depending on the kick strength $\nu$, three phases are observed.
\begin{itemize}
\item
$0<\nu<\half$: in the steady state, the cusp number is constant. This corresponds to the transparent phase of the string.
\item
$\half<\nu<1$: in the steady state, the cusp number is linear in time. The slope depends on $\nu$. This corresponds to the gray phase.
\item
$\nu>1$: cusps never reach the other side. This corresponds to the black phase.
\end{itemize}
At the $\nu=\half$ transition point, $N(t) \propto \sqrt{t}$ is observed numerically. This simple model therefore does not reproduce the quasi-logarithmic growth of cusps that we have seen in FIG. \ref{fig:number3}. It does, however, incorporate enough string dynamics to correctly distinguish between the three phases that we have seen through the cusp number function.

\section{Discussion}

The string in anti-de Sitter spacetime is one of the simplest holographic non-equilibrium systems. It is also interesting, because the induced metric on the worldsheet provides a two-dimensional toy model for gravity \cite{Dubovsky:2012wk}. In particular, the worldsheet may contain an event horizon whose formation can be studied.

This paper considered a string that connects two points on opposite sides of the boundary of global AdS$_3$. The quarks at both endpoints are kicked repeatedly (here this means that the acceleration of the string endpoint jumps at the time of kicks). This creates cusps on the string that propagate with the speed of light. The cusps eventually collide in the middle. Depending on the strength of the kicks, three qualitatively different phases were identified. In the transparent region, the cusps behave like linearized waves that pass through each other. In the black region, a worldsheet horizon forms and nothing passes through the string. In this paper, a new ``gray'' phase has also been observed. In this phase, the string length keeps growing in time. There is, however, no worldsheet horizon yet. A simple model  explained the time-evolution of the number of cusps in all three phases.

The transparent and gray regions are separated by an interesting transition point where the cusp number seems to grow approximately logarithmically. This deserves further investigations. It would be interesting to find analogous spacetime solutions in ordinary Einstein gravity.

\vspace{0.2in}   \centerline{\bf{Acknowledgments}} \vspace{0.2in}
The author would like to thank Douglas Stanford for suggesting the simple model of section IV and for his helpful comments on the manuscript.

\bibliography{global}

\end{document}